\documentclass[aps,prb,twocolumn,superscriptaddress,groupedaddress]{revtex4} 
\pdfoutput=1
\usepackage[version=3]{mhchem} % Formula subscripts using \ce{}
\usepackage[dvips]{epsfig}
\usepackage{epstopdf}
\usepackage{graphicx}  % needed for figures
\usepackage{dcolumn}   % needed for some tables
\usepackage{bm}        % for math
\usepackage{amssymb}   % for math
\usepackage{amsmath} % for math
\usepackage{color}

\usepackage[dvipsnames]{xcolor}

\begin{document}

\author{Joseph E. Losby}
\affiliation{Department of Physics, University of Alberta, Edmonton, AB, T6G 2E9, Canada}
\affiliation{National Institute for Nanotechnology, Edmonton, AB, T6G 2M9, Canada}

\author{Mark R. Freeman}
\affiliation{Department of Physics, University of Alberta, Edmonton, AB, T6G 2E9, Canada}
\affiliation{National Institute for Nanotechnology, Edmonton, AB, T6G 2M9, Canada}

\title{Spin Mechanics}

\begin{abstract}
A brief history, the current state, and future directions of \textit{spin mechanics} are presented.
\end{abstract}
\maketitle

%%% INTRO %%%
\noindent 
\subsection{SPIN MECHANICS TIMELINE}
\renewcommand{\thefootnote}{\fnsymbol{footnote}}
Electrons carry charge, magnetic moment, mass, and mechanical (spin and orbital) angular momentum.  Intensive investigation of the interplay between magnetism and charge transport, or \textit{spin electronics} has tremendously enriched our fundamental understanding of magnetic materials and enormously expanded the horizons for magnetic devices and applications.  An analogous opportunity exists in the merging of magnetism and mechanical motion, or \textit{spin mechanics}.  These include the deflection of mechanical objects due to angular momentum transfer via magnetic torques, magnetic gradient force interactions, or strains induced through magnetostrictive effects.  The idea that electrons carry a mechanical angular momentum became evident a century ago through the results of magneto-mechanical experiments seeking to determine the origin of magnetism, and mechanical implementation followed alongside the historical advancement of magnetism. 

A series of events closely tied to the development of spin mechanics is shown in the timeline, Fig. 1.  Perhaps the earliest known practical application of a ‘spin mechanical’ device by humans is through navigation thousands of years ago by use of the magnetic compass \cite{mattis}.  The discovery that the long axis of a thinned piece of lodestone always orients in the north-south direction (which we know now is due to a torque induced on the magnetization in the needle by Earth's magnetic field) had a profound effect on our history with the use of the compass in seafaring.  Predating our history, many organisms have acquired magnetoreceptive abilities that grant the ability to follow and align to small changes in Earth's field.  The simplest of them, magnetotactic bacteria, synthesize internally a linear chain of nano-sized magnetite particles that are dipolar-coupled to effectively a form a compass needle, which passively allows it to migrate with Earth's field to depths with favourable oxygenation conditions \cite{blakemore_magnetotactic}.  Below, we spotlight a few developments along the (incomplete) timeline shown in Fig. 1.  The timeline also includes key 20th century advances in magnetism having a huge bearing on the development of spin mechanics. 

\begin{figure*}[htb]
\begin{center}
\includegraphics[scale=0.95]{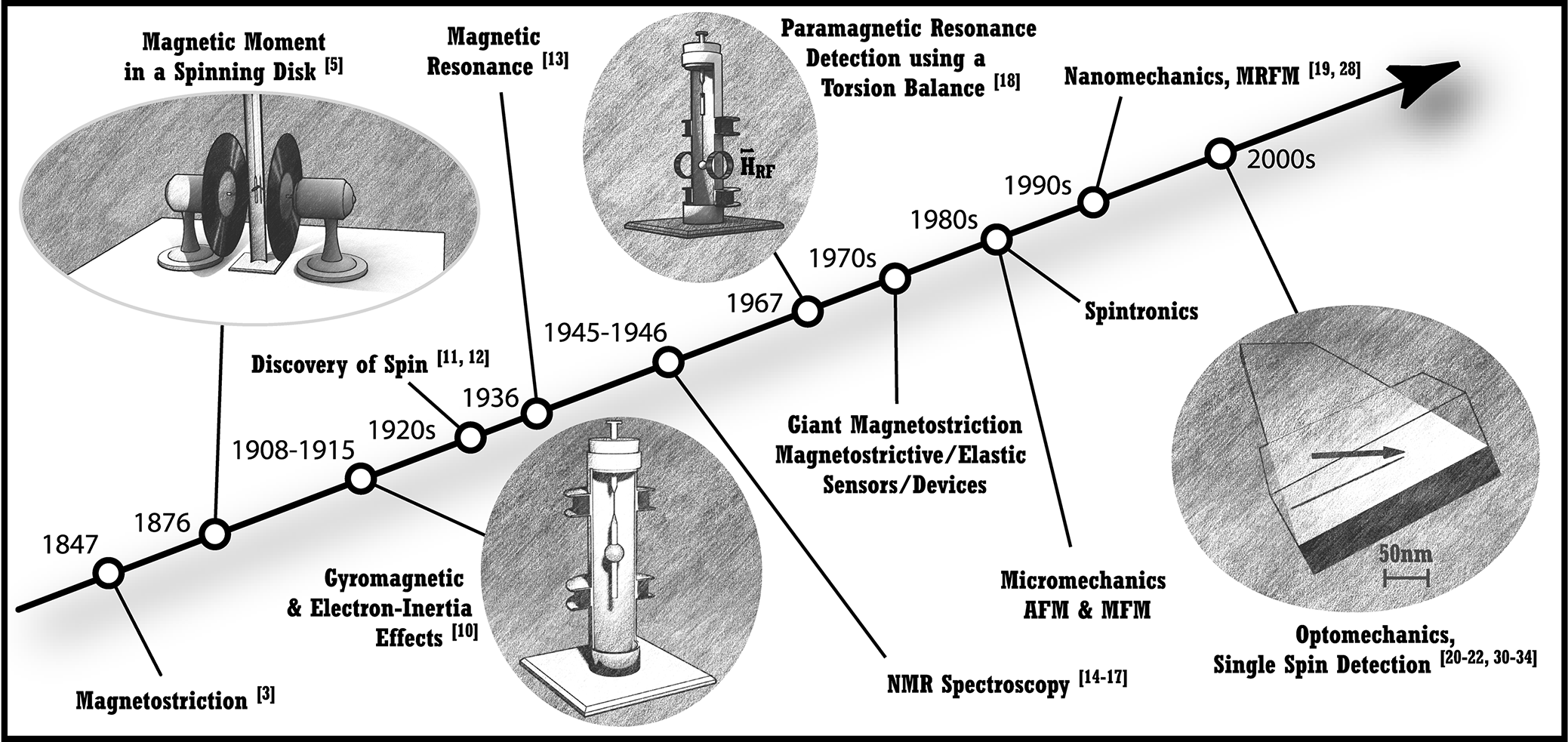}
 \caption{Timeline of key events in the development of spin mechanics, annotated with references.  Not discussed in this manuscript, due to limited space, are magnetostrictive effects and applications first discovered by Joule in 1847 \cite{jouleMS}.}
\label{timeline}
\end{center}
\end{figure*}

Amp{\`e}re, in the early 1820s, hypothesized that the magnetic field of a ferromagnetic body was produced by persistent ‘molecular currents’ \cite{ampere}.  Rowland, in the 1870s, confirmed that the mechanical rotation of electrostatically-charged disks generates magnetic fields, using an ultrasensitive torsion-fibre compass capable of measuring field changes of order 1 nT \cite{rowland}.  O.W. Richardson later suggested that a relationship should exist between the magnetization (or magnetic moment) and mechanical angular momentum in a ferromagnet (1908) \cite{richardson}.  He proposed a mechanical experiment involving a piece of iron suspended on a torsion string, which would experience a twist as the ‘magnetic atoms’ in the iron imparted a mechanical angular momentum as they switched their ‘axis’ from one direction to another along the direction of the string.  S. J. Barnett, while pondering the origin of Earth's magnetic field, suggested that a ferromagnet with no net moment should become magnetized upon mechanical rotation (inverse to Richardson’s proposal) and indicated preliminary experimental success of his theory (1909) \cite{barnett}.  He published his results in 1915 of what is now known as the Barnett effect \cite{barnett_physrev}.

In the same year, Einstein and de Haas announced their findings, believing they had confirmed the existence of the Amp{\'e}rian molecular currents \cite{einsteindehaas}.  Their experiment was similar to what was proposed by Richardson, while adding a key gain in sensitivity by changing the applied field (and switching the magnetization) at the mechanical resonance frequency of the torsion balance.  At the time, it was believed that the electron’s orbital angular momentum was solely responsible for the change in the amplitude of twist of the mechanical resonator.  The ratio of magnetic moment $\bm{m}$ to angular momentum $\bm{L}$, when calculated for a classical electron orbit or current loop, is $\bm{m}/\bm{L}=-e/2m_e$, where $m_e$ is the electron mass and $-e$ it’s charge (the ratio e/me was well known from the classic J.J. Thomson experiment), was the experimental constant sought (also known as the gyromagnetic ratio).  Einstein and de Haas’ result matched the classical gyromagnetic ratio reasonably well, within 10\% error.   

Curiously, later experiments seeking to reproduce the Richardson and Einstein-de Haas experiment all arrived at a gyromagnetic ratio of around $\bm{m}/\bm{L}=-e/m_e$, a factor of two larger than predicted \cite{barnettreview}.  This `anomaly' brought into question the accepted value.  Although the intimate association between angular momentum and ferromagnetism existed, it was not due to the persistent molecular currents.  The reason would not be known until the discovery of `the spin' and the formulation of quantum mechanics in the 1920s.  The intrinsic magnetic moment of the electron is predominantly responsible for the magnetization in ferromagnets.  

The gyromagnetic studies by Richardson, Barnett, and Einstein-de Haas, and others make up the earliest of spin mechanical measurements which preceded the development of the spin in 1925 \cite{gerlach, uhlenbeck}.  A Rev. Mod. Phys. article published by Barnett in 1935 provided a detailed summary of the classical viewpoint, and a chronology of the first century of thought and experimentation on the subject dating back to Amp{\'e}re and Weber \cite{barnettreview}. 

Magnetic resonance is at its core a spin-mechanical effect: a magnetic dipole misaligned to a local magnetic field would not precess if not for their intrinsic mechanical angular momentum (this key physical distinction relative to electric dipoles is not emphasized to physics undergraduates).  Electron spin resonance was first observed by Gorter through a calorimetric method \cite{gorter}, and then by Rabi with molecular beams \cite{rabi}.  Nuclear magnetic resonance (NMR) spectroscopy through electromagnetic induction was demonstrated by Zavoisky \cite{zavoisky}, Purcell et al. \cite{purcell}, and Bloch \cite{bloch2} independently, and laid the foundations for powerful methods now used throughout science and medicine.  Alzetta, Ascoti, Gozzini and co-workers in 1967 performed a pioneering demonstration of electron paramagnetic resonance detection using a torsion balance to record an "Einstein-de Haas torque", foreshadowing the much later resurgence of spin mechanical detection \cite{alzetta}.  The same group would revisit this work in a highly miniaturized geometry in 1996 \cite{ascoli}.        

The advancements made in semiconductor device manufacturing from the 1960s brought about the miniaturization of microelectromechanical devices.  The atomic force microscope (AFM), using a micro-cantilever with a sharp tip to probe surface forces at nanoscale resolution, was developed in the early 1980s.  Soon after, AFM tips were evolved to incorporate magnetic material.  In magnetic force microscopy (MFM), the interaction is through the magnetic gradient forces between the sample and the tip, causing a mechanical deflection of the cantilever.  Gradient force detection of magnetic resonance has been developed to sensitivity equivalent to a fraction of an electron spin, or tens of protons \cite{rugar_singlespin}.   

\subsection{MODERN SPIN MECHANICS} 
Richardson/Barnett/Einstein-de Haas effects scale-up in import for small systems with tiny moments of inertia and high angular rotation speeds.  The response of a nanoscale magnet can change qualitatively, depending upon whether the structure is anchored or free to rotate.  By 2005, Kovalev, Bauer, and collaborators, along with Chudnovsky and collaborators, were analyzing the expected consequences of the effects of rotation on magnetism in nanostructured and quantum systems \cite{kovalev, jaafar}.  Remarkable recent experiments have leapfrogged all the way to inelastic electron tunneling spectroscopy of strongly coupled spin-phonon modes in a single-molecule magnet / carbon nanotube hybrid system \cite{ganzhorn} (theory in \cite{okeefe}).  The mechanisms of angular momentum transfer between microscopic magnetic moments and their mechanical hosts are thinly understood.  Chudnovsky would note in 2004 that "the problem of spin-lattice interactions is almost as old as the quantum theory of solids" \cite{chudnovsky_ps}.  The concept of phonon angular momentum was introduced as recently as 2014 \cite{zhangl}.     

A modern Richardson/Einstein-de Haas experiment is described through Fig. 2a \cite{chudnovsky}.  Thin magnetic films (or small structures) are affixed to micro- or nanomechanical resonators while external fields induce magnetic torques that are transferred to a mechanical degree of freedom, in this case the flexural mode of a microcantilever \cite{moreland_review}. The applied AC ‘dither’ field is at the mechanical resonance frequency (frequency sweep, right side of Fig. 2a) and deflections down to the sub-nanometer can be detected using sensitive optical interferometric methods.  

\begin{figure}[htb]
\begin{center}
\includegraphics[scale=0.85]{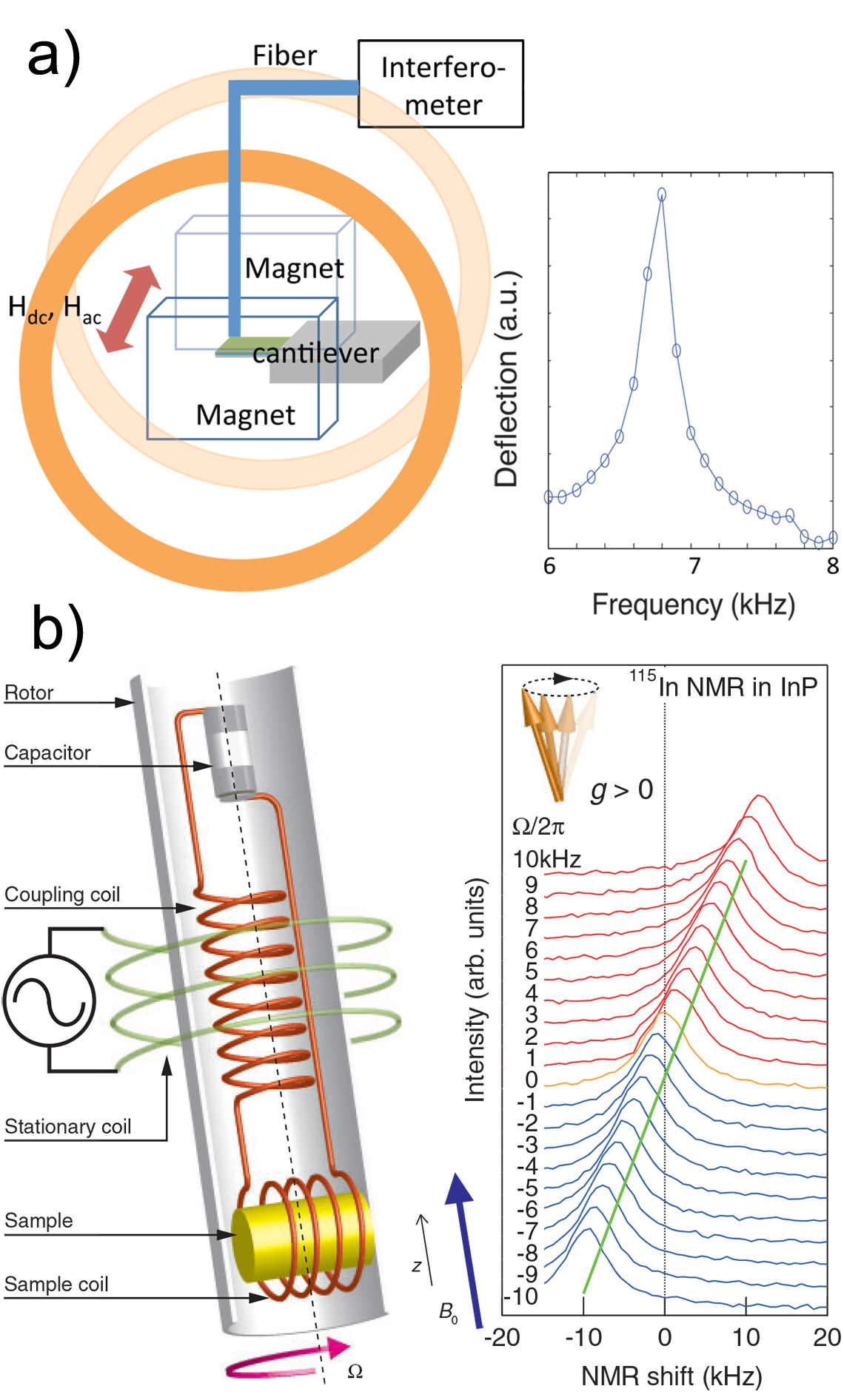}
 \caption{a) Instrumental schematic of a micromechanical Richardson/Einstein-de Haas effect driven by Hac at the fundamental flexural mode of the cantilever (frequency sweep shown on the right of the panel) while under bias by Hdc. (Ref. 27, \textcopyright 2013, The Japan Society of Applied Physics).  b) Coil-spinning method for inducing Barnett fields for readout via NMR (Ref. 29, \textcopyright 2014, IOP Publishing).  The NMR frequency shifts for the $^{115}$In nuclei in InP with angular rotation velocity is shown on the right.}
\label{timeline}
\end{center}
\end{figure}

In Fig. 2b, a schematic is shown of a recent experiment which observed Barnett fields through NMR using a ‘coil spinning’ technique \cite{chudo_nmr}.  The apparatus consists of a sample coil and a coupling coil, both spinning within a stationary coil connected to a NMR spectrometer.  Through mutual induction the RF field from the stationary coil is received by the coupling coil and transferred to the sample coil, inducing the NMR signal.  The Barnett field is proportional to the angular frequency Ω, and modifies the applied field $\bf{B}_0$, resulting in NMR frequency shifts.  This is shown in the resonance spectra for $^{115}$In nuclei under various sample rotation speeds (Fig 2b, right).  

\begin{figure}[htb]
\begin{center}
\includegraphics[scale=0.85]{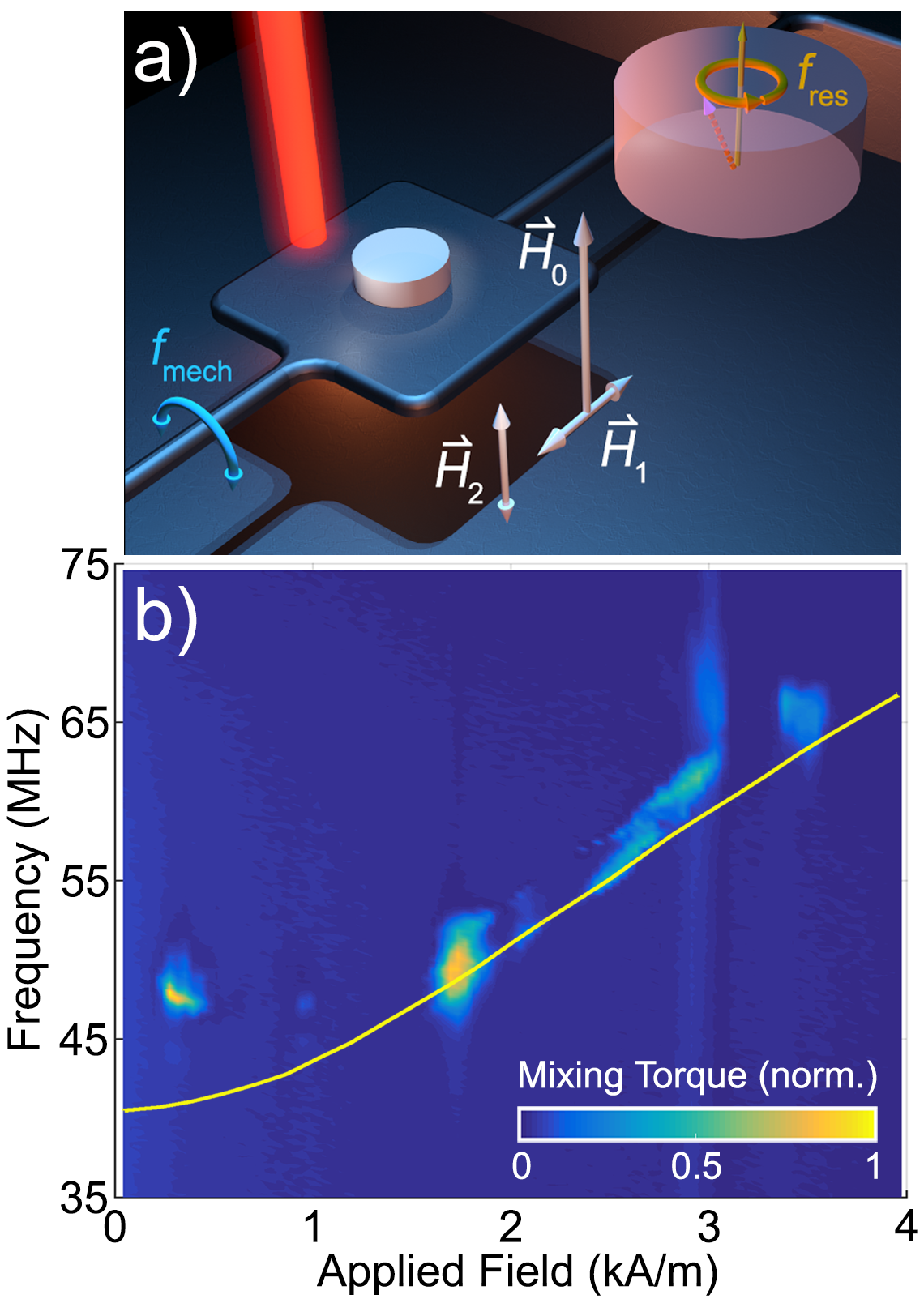}
 \caption{a) Applied field geometry for torque-mixing magnetic resonance spectroscopy through mechanical detection of a nanoscale torsional resonator. b) Mixing torque spectroscopy of a permalloy disk with a vortex magnetization texture.  The vortex gyrotropic resonance mode shows interactions with nanoscale surface imperfections with applied field.  The solid line is from simulation results using a pristine disk.}
\label{timeline}
\end{center}
\end{figure}

A recently developed torque method for magnetic resonance spectroscopy is shown in Fig. 3a \cite{LosbyTMRS}.  The method permits direct detection of the transverse component of precessing dipole moments (inset), and parallels the conventional approach to NMR, where this is detected inductively.  Here, a mesoscopic magnetic sample is attached to a nanomechanical torsional resonator and placed under bias by the static field $\bf{H}_0$.  The RF field $\bf{H}_1$ (at frequency $f_1$) drives the precession of the moments at $f_{res}$.  An additional RF field $\bf{H}_2$ (at $f_2$) cooperates with $\bf{H}_1$ to generate sum and difference frequency ‘torque-mixing’ components proportional to the magnetic resonance amplitude.  By applying $f_2$ and $f_1$ such that their difference is the mechanical mode fmech of the torsional resonator, the magnetic resonance can be read out with high sensitivity.  Spectroscopy is performed by sweeping $f_2$ and $f_1$ together while maintaining the $f_{mech}$ difference.   

An example of torque-mixing resonance spectroscopy is shown in Fig. 3b for a Permalloy disk (Ni$_{80}$Fe$_{20}$, 15 nm thick and 2 µm in diameter).  Such a structure holds a low-field vortex magnetization state, with a core pointing out of plane to the disk surface.  The lowest order magnetic resonance mode of the vortex texture is that of a precession of the core about an equilibrium.   With applied field, the equilibrium of the core is ‘pushed’ towards the edge of the disk and its precession frequency is blue-shifted, as seen in the figure.  The core, with a high exchange energy density, can probe the magnetic landscape.  The ‘dropouts’ seen in the evolution of the magnetic resonance signal with applied field are due to pinning events as the core interacts with nanoscale grain boundaries inherent in Permalloy, and also observable as Barkhausen transitions in the net magnetization \cite{burgess, fanisani}.  The micromagnetic simulation results for a ‘pristine’ disk is shown overlaid. 

\subsection {CAVITY TORSIONAL OPTOMECHANICS}
Most of the spin-mechanical phase space between millions-of-Bohr-magneton objects and single spin systems remains unexplored but is now accessible, owing to advances in related, enabling technologies.  Experimental capabilities for detecting nanomechanical motion have been revolutionized by the development of cavity optomechanics.  In these systems a micro- or nanoscale mechanical resonator is embedded in a high finesse optical cavity.  A dispersive coupling of the mechanical modulation with the cavity results in an optical resonance frequency shift, which is detected with extremely high sensitivity. Displacements of a nanostructure corresponding to a small fraction of the diameter of a proton have been measured.  
\begin{figure}[htb]
\begin{center}
\includegraphics[scale=0.65]{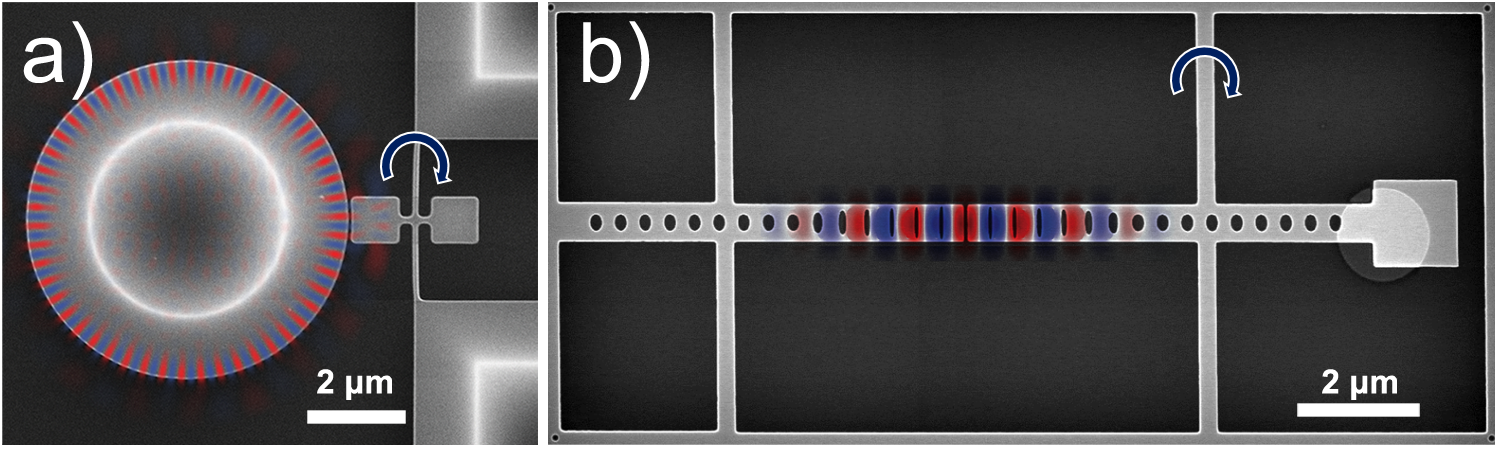}
 \caption{a) Scanning electron micrograph of a nanoscale torsional optomechanical system (from Ref. 33). A mechanical resonator operating at the fundamental torsion mode is coupled to the whispering gallery optical mode (shown overlaid with the optical mode profile) evanescently (Reproduced with permission from P.H. Kim et al., Appl. Phys. Lett., 102, 053102. \textcopyright 2013, AIP Publishing LLC).  b) A photonic crystal cavity optomechanical torque magnetometer consisting of a split-beam geometry of mirrors (Ref. 34).  The suspended mirror holds a mesoscopic Permalloy element 40 nm thick and operates at a torsional mode, dispersively modifying the optical resonance.  The normalized field distribution (Ey) the optical mode is shown overlaid with the cavity.}
\label{timeline}
\end{center}
\end{figure}

A significant advance in torsional nanomechanics has been achieved through coupling to an optical whispering gallery mode in a silicon microdisk, as presented in Fig. 4a \cite{pkim}.  Light is coupled into the microdisk using a single-mode tapered fiber. The optical field profile from simulation is overlaid on the scanning electron micrograph.  The motion of the torsional resonator interacts with evanescent fields, modulating the effective index of refraction (and thus resonance frequency) of the optical mode.  Mechanical deflections corresponding to torques down to the 10-20 Nm scale have been reported.  The earliest implementation of a photonic crystal cavity optomechanical torque magnetometer is shown in Fig. 4b \cite{wu_notm}.  In this scheme, a magnetic element is placed at the end of a suspended structure serving as an optical ‘mirror’, which can operate mechanically at the torsional mode.  The suspended mirror is optically coupled to an anchored mirror receiving light from an optical fiber.  To minimize radiation losses, the periodic holes defined in the structures are tapered to the dimension of the gap between the two mirrors (optical field profile shown below).  With an applied AC magnetic field, the resonating mechanical structure dispersively causes a frequency shift of the optical mode.  

\subsection{CONCLUSION}
The magnetism sub-discipline of spin mechanics is at an exciting stage.  Direct experimental insights on the behaviour of spin-rotation coupling in a wide variety of materials is key to a fuller basic understanding of magnetism.  Resonant detection of spin angular momentum opens the door to physics not yet explored, such as the timescales associated with the Richardson/Einstein-de Haas effect.  The coherent coupling of spin and motion potentially leads to mechanical control of magnetism for applications. Numerous other benefits of spin mechanics will be powerful new mechanical tools for the experimental magnetician's kit, complementary to existing methods, including fully broadband optomechanical labs-on-a-chip for analysis (magnetometry and resonance spectroscopy) of structures from magnetic nanodevices to nanoparticles. 

The fourth international workshop on Spin Mechanics will be held on February 20-25, 2017 in Lake Louise, Alberta.

%%% ACKNOWLEDGEMENTS %%%
\section*{ACKNOWLEDGEMENTS}
The authors very gratefully acknowledge support from Natural Science and Engineering Research Council, Canada Research Chairs, Alberta Innovates Technology Futures, and National Institute for Nanotechnology.  The authors also thank David Fortin for the timeline design and 3d renderings (Fig. 1 and Fig 3a), Fatemeh Fani Sani for micromagnetic simulation (Fig. 3b), and the respective authors for contributing material used in this manuscript

\newpage

%%%% BIBLIOGRAPHY %%%%
\bibliography{Pic}
\end{document}